\documentclass[10pt, a4paper]{article}
\usepackage[utf8]{inputenc}
\usepackage{fullpage}
\usepackage{graphicx}
\usepackage[usenames,dvipsnames]{color}
\usepackage{cite}
\usepackage{amsmath}
\usepackage{amssymb}
\usepackage{multirow}
\usepackage{booktabs}
\usepackage{pbox}
\usepackage{tabularx}
\usepackage{multimedia}
\usepackage{float}
\usepackage{newfloat}
\usepackage{ccaption}
\usepackage{authblk}

\DeclareFloatingEnvironment[fileext=lop]{video}

\newcommand{\supfigA}{Supplementary Video~S1}
\newcommand{\supfigB}{Supplementary Video~S2}
\newcommand{\supfigC}{Supplementary Figure~S3}
\newcommand{\supfigD}{Supplementary Figure~S4}

\newcommand{\supfigF}{Supplementary Figure~S6}
\newcommand{\supfigG}{Supplementary Figure~S7}

\newcommand{\supfigI}{Supplementary Figure~S9}

\begin{document}

  \title{Early processing of consonance and dissonance in human auditory cortex}

  \author[12*]{Alejandro Tabas}
  \author[3*]{Martin Andermann}
  \author[3]{Valeria Sebold}
  \author[3]{Helmut Riedel}
  \author[14+]{Emili Balaguer-Ballester}
  \author[3+]{Andr\'{e} Rupp}

  \affil[1]{Faculty of Science and Technology, Bournemouth University; UK}
  \affil[2]{Max Planck Institute for Human Cognitive and Brain Sciences; Leipzig, Germany}
  \affil[3]{Biomanetism Section, Heidelberg University; Germany}
  \affil[4]{Bernstein Center for Computational Neuroscience Heidelberg/Mannheim; Germany}
  \affil[*]{Joint first author}
  \affil[+]{Joint last author}

  \date{}

\maketitle

\begin{abstract}
  Pitch is the perceptual correlate of sound’s periodicity and a fundamental property of the auditory sensation. 
  The interaction of two or more pitches gives rise to a sensation that can be characterized by its degree of consonance or dissonance, an essential element in the perception of highly complex sound patterns. In the current study, we investigated the neuromagnetic representations of consonant and dissonant musical dyads using a new model of cortical activity, in an effort to assess the possible involvement of pitch-specific neural mechanisms in consonance processing at early cortical stages.
    
  In the first step of the study, we developed a novel model of cortical pitch processing designed to explain the morphology of the pitch onset response (POR), a pitch-specific subcomponent of the auditory evoked N100 component in the human auditory cortex. The model explains the neural mechanisms underlying the generation of the POR and quantitatively accounts for the relation between its peak latency and the perceived pitch. 

  Furthermore, we applied magnetoencephalography (MEG) to record the POR as elicited by six consonant and dissonant dyads. The peak latency of the POR was strongly modulated by the degree of consonance within the stimuli; specifically, the most dissonant dyad exhibited a POR with a latency that was $\sim$30\,ms longer than that of the most consonant dyad, an effect that greatly exceeds the expected latency difference induced by a single pitch sound. 

  Our model was able to predict the POR latency pattern observed in the neuromagnetic data, and to generalize this prediction to additional dyads that were not included in the MEG experiment. These results indicate that the neural mechanisms responsible for pitch processing exhibit an intrinsic differential response to concurrent consonant and dissonant pitch combinations, suggesting that the perception of consonance and dissonance might be an emergent property of the pitch processing system in human auditory cortex.
\end{abstract}

\section{Introduction}

  Pitch is the perceptual correlate of the periodicity in a sound's waveform. It is a fundamental attribute of auditory sensation that forms the basis of both music and speech perception; consequently, understanding the neural foundations of pitch is a major challenge in auditory neuroscience.

  A combination of two sounds that simultaneously elicits two different pitches is referred to as a dyad, and the pitch interactions within a dyad give rise to a sensation that can be characterized by its \emph{consonance} or \emph{dissonance}. Loudness, timbre, and the absolute fundamental frequencies of the sounds can have subtle effects on whether a dyad is perceived as consonant or dissonant; however, the dominant factor in determining the degree of consonance is the \emph{relationship} between the fundamental periods of the sounds that make up the dyad \cite{Kameoka1969a}: Simple periodicity ratios result in more consonant sensations; on the other hand, sensation becomes more and more dissonant as the complexity of the periodicity ratio increases \cite{Helmholtz1863, Krueger1913, Plomp1965}. Previously, it was argued that dissonance correlates with the beating, or \emph{roughness} sensation that is elicited by the interfering regularities of the involved sounds \cite{Helmholtz1863, Krueger1913, Plomp1965}. However, listeners that showed impaired pitch perception but were sensitive to beating and roughness have been reported to be unable to differentiate between consonant and dissonant dyads \cite{Cousineau2012, Tramo2006}. This suggests that pitch- rather than roughness-related auditory processing is responsible for the emergence of consonance.

  Neurophysiological evidence for a tight link between consonance and pitch has recently been provided by Bidelman and colleagues \cite{Bidelman2014}; they showed, using electroencephalography (EEG), that the amplitude of the cortical pitch onset response (POR) is strongly modulated by a dyad's perceived consonance. The POR is a pitch-selective component of the transient auditory evoked potential/field (AEP/AEF) that occurs within the time range of the well-known N100 wave \cite{Naatanen1987, Alain1997}, around 100 ms after pitch onset. The morphology of the POR is strongly correlated with the perceived pitch in single tones: its latency scales linearly with the period of the sound, and its amplitude increases with the strength of the pitch percept \cite{Krumbholz2003, Ritter2005, Tabas2016}. The neural sources of the POR are located in the anterolateral section of Heschl's gyrus (alHG) in auditory cortex \cite{Krumbholz2003, Ritter2005, Bidelman2014}, consistently with the anatomical location of pitch-selective neurons in non-human primates (e.g.,~\cite{Bendor2006, Bendor2010, Bizley2013, Feng2017}), and with pitch-selective regions that were reported for human listeners \cite{Griffiths2001, Penagos2004, Brugge2009, Norman2013, Moerel2012}. Further experiments in humans have demonstrated that the dyad-evoked frequency-following response in the brainstem is predictive for the perceived consonance of a dyad \cite{Bidelman2009, Bidelman2011, Bidelman2013}; also, functional magnetic resonance imaging (fMRI) studies showing selective activation to consonance/dissonance contrasts in the superior temporal gyrus \cite{Peretz2001} and in frontal cortex \cite{Minati2009} lead the auditory community to link neural representations of consonance and dissonance with higher, cognitive processes \cite{Seger2013}.
 
  In this study, we used a combined experimental and theoretical approach to assess whether consonance and pitch share similar processing mechanisms in human auditory cortex. Towards this goal, we first developed an innovative, realistic model of cortical ensemble responses to pitch, specifically designed to understand the mesoscopic representation of pitch in alHG. The model can account, mechanistically, for the latency effects in the POR that have been robustly reported for decades in multiple experimental settings \cite{Krumbholz2003, Ritter2005, Gutschalk2004} but, up to now, remained poorly understood. Second, we recorded the AEF elicited by consonant and dissonant dyads using magnetoencephalography (MEG); here, our experimental results revealed a strong correlation between POR latency and the degree of consonance, extending previous EEG findings \cite{Bidelman2014}. Finally, we aimed to replicate the results from our MEG experiment using our model. If the hypothesis that consonance and pitch are processed by similar mechanisms in cortex is correct, we would expect the model to explain the dependence of POR latency on the degree of consonance \emph{without} the inclusion of higher processing stages within the auditory hierarchy \cite{Peretz2001, Minati2009}. In line with this hypothesis, the model provided a quantitative, mechanistic explanation for the relationship between POR dynamics and consonance, suggesting that consonance and dissonance perception might be linked to the pitch processing sub-regions of the auditory cortex.

\section{Results}

  \subsection{Neural mechanisms underlying pitch processing in auditory cortex}

    \paragraph{Model overview}

      We introduced a model of cortical pitch processing designed to explain the morphology of the POR as elicited by the onset of consonant and dissonant dyads (see full description in Methods). The model consists of three processing stages located at different levels of the auditory hierarchy. In the first stage, an array of idealized coincidence detector units extracts periodicities from the auditory nerve activity in response to the target stimulus \cite{Meddis2006, Balaguer2008}. Subsequently, the second and third stages, putatively located at adjacent locations within alHG, generate a stable pitch characterization from the first stage, previous representation.

\begin{figure}
        \centering
        \includegraphics[width=1\textwidth]{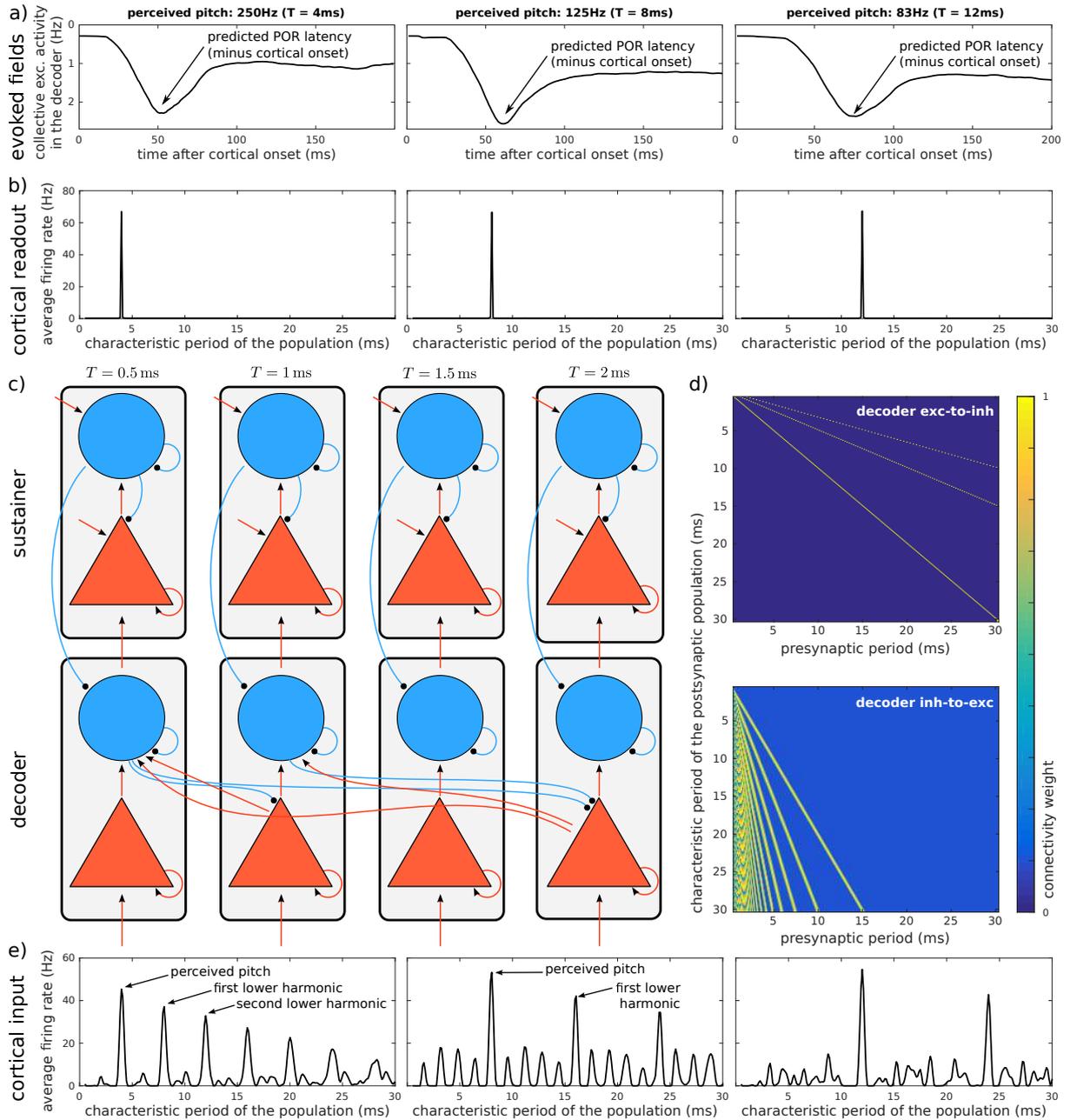}
        \caption{\textbf{Basic schematics of the model.}
        a) Average response across populations of the excitatory ensembles of the decoder network, accounting for the evoked fields in the cortical sources of the POR, in response to three stimuli with different pitches.
        Stimuli were iterated rippled noises (16 iterations, bandpass filtered between 0.8 and 3.2\,kHz) with pitches: 250\,Hz, 125\,Hz, and 83\,Hz. 
        b) Excitatory activation, averaged between 150 and 200\,ms after sound onset, in the different ensembles of the decoder network elicited by the same stimuli as above.
        c) Model architecture.
        d) Connectivity weights between excitatory and inhibitory ensembles in the decoder network.
        e) Typical input to the decoder corresponding to the three IRN stimuli used above, averaged between 150 and 200\,ms after sound onset (see also \supfigA).}
        \label{fig:mod:diagram}
      \end{figure}

      Auditory nerve responses were generated by a recent model of the auditory periphery \cite{Zilany2014, Zilany2009}; periodicity detection was implemented using the principles of the autocorrelation models of pitch \cite{Licklider1951, Meddis1997, Meddis2006, Balaguer2008}. At this stage, the representation of the stimulus shows a harmonic shape along the periodicity axis, with prominent peaks of activation at the neurons which encode the pitch of the stimulus and its lower harmonics (see Figure~\ref{fig:mod:diagram}e).

      The array of periodicity detectors provides excitatory input to a first cortical processing stage, termed the \emph{decoder} network in this study. The \emph{decoder} network is putatively located in alHG and effectively extracts the pitch value(s) from the subcortical input. The decoder network connects to a second cortical ensembles network, termed \emph{sustainer}; this stage integrates the decoder network output and modulates it through cortico-cortical top-down efferents. The \emph{sustainer} reinforces and sustains the decoder mechanism, reminiscent to recent models of perceptual decision making \cite{Wimmer2015}.

      Both decoder and sustainer comprise a network of cortical microcolumns, each of which is tuned to a specific pitch value along the human perceptual range (see Methods for details). Pitch is encoded in the active pitch-selective populations of the processing network (see Figure~\ref{fig:mod:diagram}b), in agreement with cortical recordings in non-human mammals \cite{Wang2012, Bizley2013, Gao2016}.
         
\begin{figure}[b!]
        \centering
        \includegraphics[width=1\textwidth]{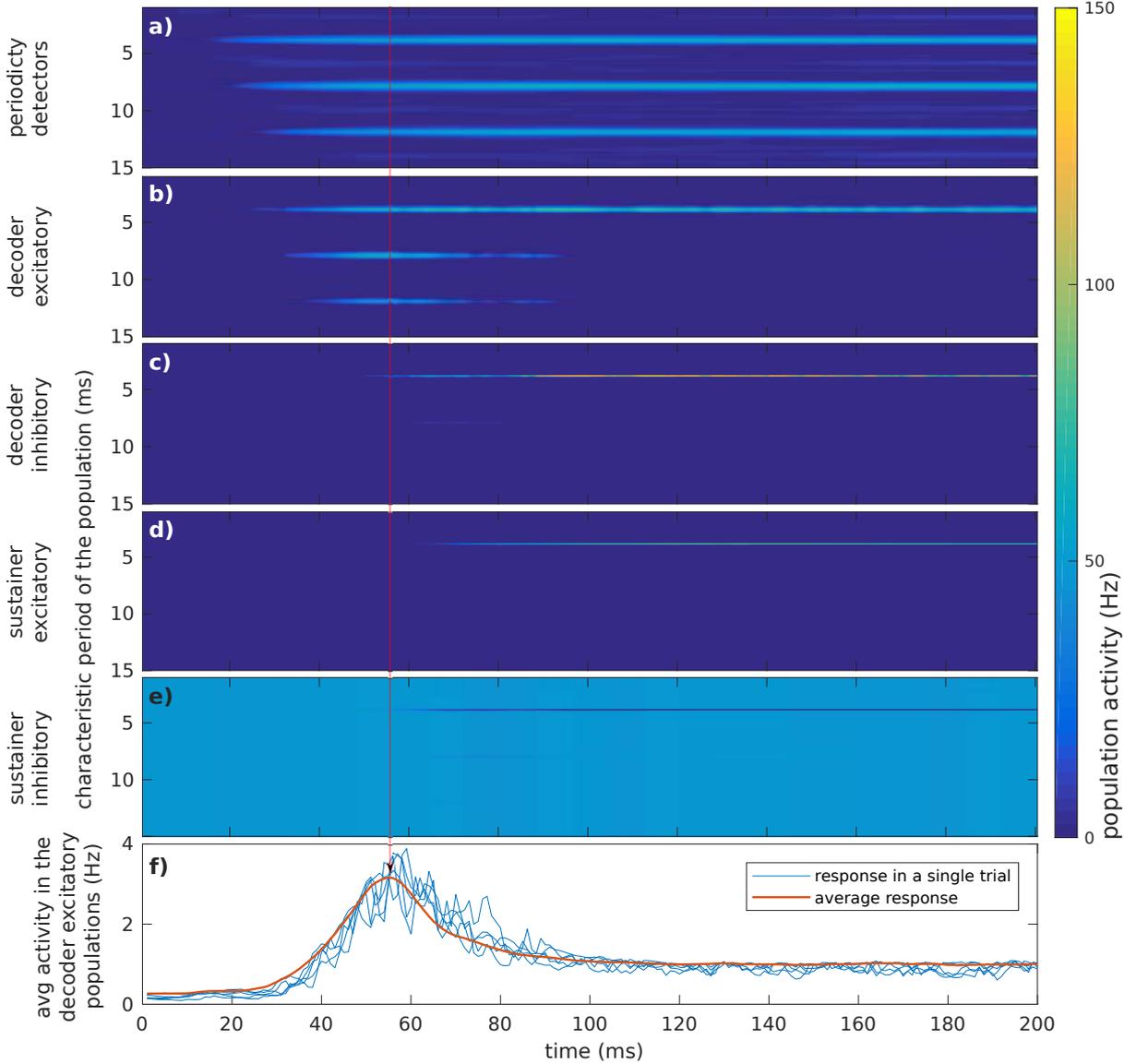}
        \caption{\textbf{Illustration of the decoding process.} The plots show the evolution of key variables of the model during the processing of the first 200\,ms of an iterated rippled noise with a fundamental period of $T = 4\,$ms (parameters were as in Figure~\ref{fig:mod:diagram}). a)--e) Evolution of the model neural ensembles encoding characteristic periods between 0.5\,ms and 15\,ms. a) Activity of periodicity detectors within the first stage of the model. b) and c) Activity of excitatory and inhibitory ensembles in the decoder network. d) and e) Activity of excitatory and inhibitory activities in the sustainer network. f) Aggregated excitatory activity in the decoder. Detailed dynamics of the process are illustrated in an animation in \supfigA.}
        \label{fig:mod:dec}
      \end{figure}

      Microcolumns in the cortical networks are modelled as blocks comprising an excitatory and inhibitory neural ensemble (see Figure~\ref{fig:mod:diagram}c). Ensembles communicate with each other through realistic synapses. Connectivity weights between populations in the decoder network (see Figure~\ref{fig:mod:diagram}e) are specifically tuned to facilitate the inhibition of the lower harmonics elicited in the periodicity detectors (see Figure~\ref{fig:mod:dec}; a video detailing the integration process is available in \supfigA). Similar connectivity patterns have been consistently found in the mammalian auditory cortex (see~\cite{Wang2013} for a review); moreover, neurons mapping harmonic templates to a pitch-selective representation like those introduced in this model have been recently reported in the primate auditory cortex \cite{Feng2017}.

\begin{figure}
        \centering
        \includegraphics[width=1\textwidth]{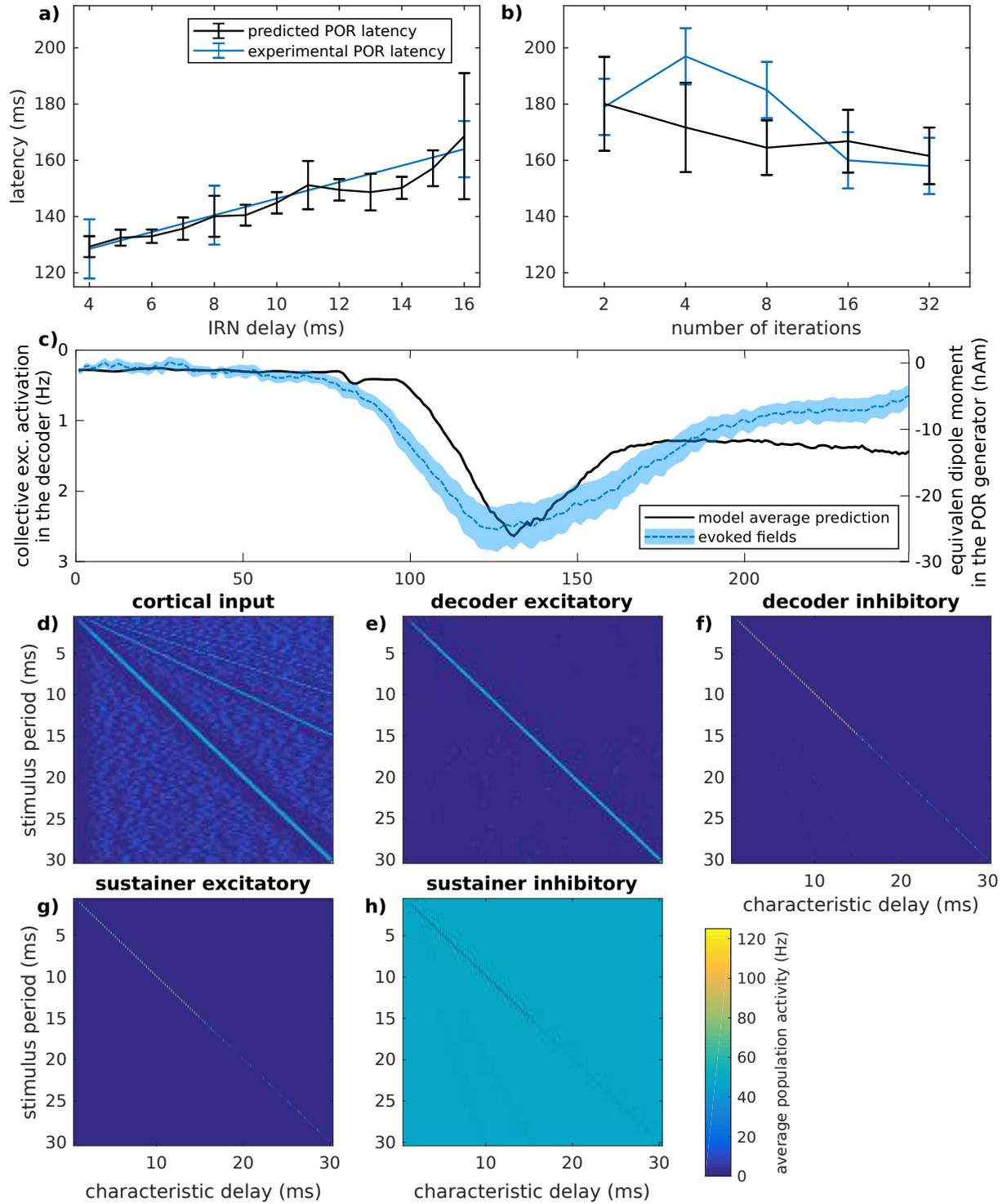}
        \caption{\textbf{Model predictions for iterated rippled noise pitch.} a)--b) Latency predictions for iterated rippled noise compared with experimental data obtained by a previous study \cite{Krumbholz2003} for the same stimuli. 
        c) Comparison of the collective activation of the excitatory ensembles in the decoder (computed as an average across populations) with the equivalent dipole moment elicited at the generator of the POR; stimulus was an IRN of 16 iterations and a delay of 8\,ms.
        d)--h) Averaged responses at: (d) periodicity detectors, (e/f) excitatory/inhibitory ensembles in the decoder, (g/h) excitatory/inhibitory ensembles in the sustainer.}
        \label{fig:mod:IRNs}
      \end{figure}

      The detailed formulation of the model allows us to perform quantitative predictions regarding the neuromagnetic field that are elicited by the activation of each of the cortical networks within the model (Figure~\ref{fig:mod:diagram}b). More specifically, the equivalent dipole moment elicited by each of the networks is monotonically related to the aggregated excitatory activation of its ensembles \cite{Kiebel2008} (see Methods for details), while the characteristic period of the excitatory population with the largest activity in the network corresponds to the perceived pitch (see Figures~\ref{fig:mod:IRNs}b and~c). Below, we will argue that the characteristic responses of the decoder network during pitch processing can be identified with the responses of the neural sources of the POR. 

    \paragraph{Dynamics of the decoder network}

      Figure~\ref{fig:mod:dec} illustrates an example of the model dynamics in response to a stimulus with a pitch corresponding to $T = 4\,$ms (i.e., $f = 250\,$Hz; details are shown in \supfigA). In a first step, periodicity detectors, tuned to $T \simeq 4\,$ms, become active after $t_1 \sim 1.25\,T$ \cite{Wiegrebe2001} (see the top prominent horizontal line at $T = 4$\,ms in Figure~\ref{fig:mod:dec}a). These active neurons provide the bottom-up excitatory input to the excitatory ensemble in the corresponding decoder network column (see Figure~\ref{fig:mod:dec}b). Likewise, the harmonics of the stimulus' period (i.e., $2\,T$, $3\,T$ etc.) are subsequently represented in the subcortical model after $t_2 = 2\,t_1$, $t_3 = 3\,t_1$ etc., and provide the input to the corresponding excitatory populations in the decoder network (see Figure~\ref{fig:mod:dec}b).

      Excitatory ensembles characterized by the periods of the harmonic series $\{T, 2\,T 3\,T\, \dots\}$ are connected to inhibitory neural populations identified by the fundamental period $T$ (see Figure~\ref{fig:mod:diagram}). Synaptic efficacy is tuned such that the inhibitory drive is strong enough to elicit activity only when a sufficient number of excitatory inputs (typically 3) are simultaneously active, thus providing robustness to the upcoming decoding process. Thus, the inhibitory population characterized by $T \simeq 4$\,ms becomes active only when it receives simultaneous synaptic drive from the excitatory ensembles characterizing the periods $T = 4\,$ms, $2T = 8\,$ms, and $3T = 12\,$ms (see Figure~\ref{fig:mod:dec}C).

      Correspondingly, the inhibitory ensemble associated with the period $T$ is connected to the excitatory populations encoding the lower harmonics $\{2\,T, 3\,T, 4\,T, \dots\}$ (see Figure~\ref{fig:mod:diagram}). Thus, when active, the inhibitory activity progressively silences populations that are activated by the periodicity detectors but that do not correspond to the fundamental period of the stimulus (see the shunting process in the decoder excitatory network between 60\,ms and 100\,ms in Figure~\ref{fig:mod:dec}b).

      The processing dynamics described above explains the neural mechanisms underlying the morphology of the POR. The aggregated excitatory activity in the decoder, monotonically related to the neuromagnetic fields elicited by this network, is shown in Figure~\ref{fig:mod:dec}f. The accumulation of excitatory activity corresponding to the harmonic series of the period of the stimulus $T$ results in the increase of the simulated field magnitude observed between 10\,ms and 65\,ms in Figure~\ref{fig:mod:dec}f. The subsequent decay of the model collective excitatory response between 70\,ms and 120\,ms in the figure, is similarly caused by the action of the most activated inhibitory ensemble on the excitatory populations encoding the lower harmonics. We identify the maximum in the aggregated excitatory activity, corresponding to the time point in which the model performs a perceptual decision about the pitch of the stimulus, with the POR latency (further details regarding this correspondence are shown in \supfigA).

      The proposed mechanism also explains, quantitatively, the dependence of the POR latency with the period of the stimulus. A periodicity detector tuned to $T$ needs $\gtrsim1.25\,T$ to robustly detect a periodicity $T$ in the auditory nerve activity \cite{Wiegrebe2001}. Thus, the arrival of sufficient excitatory drive to activate an inhibitory ensemble shows a dependence with several periods of the stimulus. This result also provides a mechanistic explanation for the minimum stimulus duration required for robust pitch discrimination, which is around four times the period of the stimulus \cite{Krumbholz2003}.

    \paragraph{Dynamics of the sustainer network}

      The dynamics of the decoder network are sufficient to explain how the harmonic representations held in the first step of the model are transformed into the final representations shown in Figures~\ref{fig:mod:diagram}b and~\ref{fig:mod:diagram}c. However, after the transformation has taken place, the excitatory ensembles  corresponding to the lower harmonics of the stimulus' pitch are no longer active, and hence the inhibitory population silencing them loses its drive. Thus, without top-down control, the decoder network would rapidly reset and need to repeatedly extract the pitch from the harmonic representation, eliciting a series of PORs; this, however, does not reflect the experimental observations (Figure~\ref{fig:mod:diagram}a). The role of the sustainer network is to regulate the dynamics of the decoder network once the pitch value has been extracted from the harmonic representation, in order to effectively \emph{sustain} the perceptual decision until a significant change is produced in the cortical input.

      In the absence of external input, the sustainer network rests at equilibrium, with a steady activation in the inhibitory populations and complete deactivation of the excitatory populations (see Figures~\ref{fig:mod:dec}d and~e).

      Excitatory/inhibitory ensembles in the sustainer receive direct input from their respective excitatory/inhibitory counterparts in the decoder (see Figure~\ref{fig:mod:diagram}d). Thus, a significantly active inhibitory population in the decoder effectively silences the steady activity of the analogous inhibitory population in the sustainer. If this afferent drive coincides with a strong activation of the corresponding excitatory population in the decoder, the combined bottom-up input results in a strong activation in the equivalent excitatory population within the sustainer (see Figures~\ref{fig:mod:dec}d and~\ref{fig:mod:dec}e).

      Top-down efferents connect each excitatory population in the sustainer with its inhibitory counterpart in the decoder network (see~\ref{fig:mod:diagram}b), compensating for the loss of excitatory drive of the silenced populations for as long as the subcortical input remains unchanged (the behaviour of the network under pitch changes is described in \supfigI). 

    \paragraph{Model predictions}

      The POR is defined as the subcomponent of the N100 transient that responds selectively to pitch onset and pitch changes \cite{Krumbholz2003, Ritter2005, Seither2006}. In order to isolate the POR from other subcomponents of the N100 like, e.g., the energy onset response (EOR), experimental setups use iterated rippled noise (IRN) preceded by a noise burst of the same energy and bandwidth \cite{Krumbholz2003, Ritter2005, Bidelman2014}; the POR is then measured as the transient elicited at the transition between noise and IRN (i.e., at the pitch onset). Thus, we tested the predictive power of our model using IRN stimuli with different pitch values (an isolated POR can be elicited using energy-balanced stimuli, but see~\supfigF for an example of how more general predictions could be drawn for other stimuli.)

      Latency predictions of the POR elicited by IRN stimuli are compared with experimental data in Figures~\ref{fig:mod:IRNs}a and~\ref{fig:mod:IRNs}b. Results show that the model replicates the relation between the POR latency and the period of the stimuli as typically reported in the MEG literature \cite{Krumbholz2003, Ritter2005}. Replication of these results for IRN stimuli with different parametrisations are shown in \supfigD. 

      To ensure that the model is correctly representing the pitch of the IRN stimuli used in the experiment, we also plotted the average activation in the different ensembles in Figures~\ref{fig:mod:IRNs}d--h. Since periodicity detectors and excitatory neurons at the decoder do not show selective activation with pitch (e.g., neurons representing harmonics of the actual pitch value also activate during the decoding process), the perceptual readout of the model can only be robustly measured in the inhibitory populations in the decoder and the populations in the sustainer. Perceptual results in Figures~\ref{fig:mod:IRNs} and \supfigD indicate that the model is able to robustly represent the pitch of the stimulus when at least two harmonics are present in the cortical input (since we only consider periodicity detectors tuned to periods under $T_{\max} = 30$\,ms, the highest period robustly extracted by the model is $T_{\max} / 2 = 15$\,ms; a larger pitch range could be easily achieved by increasing $T_{\max}$). Robust pitch extraction is shown for IRN stimuli with different parametrisations, pure tones, harmonic complex tones and click trains in \supfigC.

  \subsection{Neuromagnetic correlates of consonance and dissonance in auditory cortex}

\begin{figure}[htb!]
      \centering
      \includegraphics[width=1\textwidth]{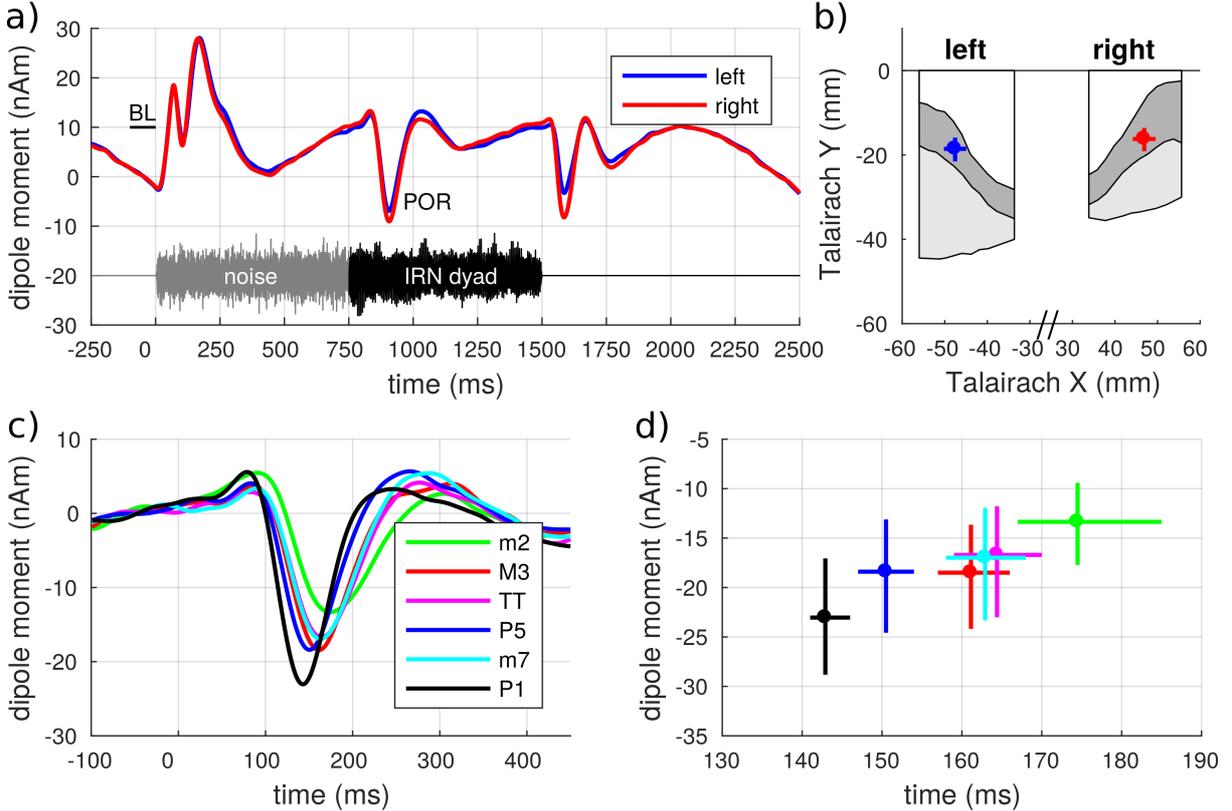}
      \caption{\textbf{Auditory fields evoked at dyad onset.} a) MEG grand-mean source waveforms in response to the pooled stimulus conditions. The course of the stimuli is shown in grey (noise) and black (IRN) below the source waveforms; note the prominent negative POR deflection (N1m) at the transition from the first to the second stimulus segment. BL = baseline. b) Projection of the dipole locations (means and 99,\% Bootstrap confidence intervals) onto the axial view of auditory cortex as suggested by Leonard et al. \cite{Leonard1998}. c) Morphology of the POR in response to the dyad onset in the single experimental conditions (second stimulus segment), pooled over hemispheres. d) 99\,\% Bootstrap confidence intervals for the POR amplitudes and latencies in the single experimental conditions. In subplots, b) and d), confidence intervals are bias-corrected and accelerated, as recommended by Efron and Tibshirani \cite{Efron1993}.}
      \label{fig:aefs}
    \end{figure}

    Next, we recorded neuromagnetic fields evoked by six different dyads from 37 normal hearing subjects. Data were preprocessed using standard MEG procedures, and equivalent current dipoles were fitted, independently, for the POR for each subject and hemisphere, pooled over conditions (see Methods). Dipole locations in Talairach space are plotted in Figure~\ref{fig:aefs}b.

    Dyads consisted of two IRN sounds. The lower note pitch was 160\,Hz; the pitch of the upper note was adjusted accordingly to form either a consonant dyad (unison, P1; perfect fifth, P5; major third, M3) or a dissonant dyad (tritone, TT; minor seventh, m7; minor second, m2). To dissociate the EOR in Planum temporale from the POR in alHG, the dyads were preceded by an energy-balanced noise segment, cross-faded with the dyad to avoid discontinuous waveforms (like for the single IRN sounds analyzed in the previous section; see Methods).

    Figure~\ref{fig:aefs}A presents the MEG grand-mean source waveforms, for both hemispheres, in response to the six stimulus conditions. The noise onset from silence (depicted in grey below the source waveforms) was followed by a transient P1m-N1m-P2m AEF complex. Since the first stimulus segment did not vary between conditions, we did not expect to find any significant differences in the corresponding neuromagnetic activity at this point.

    In contrast, the transition to the second stimulus segment (IRN dyads;  black signal below the source waveforms) elicited prominent POR waves, and the morphology of the POR varied considerably between conditions. Figure~\ref{fig:aefs}C show close-up views of the POR. Consonant dyads (pooled conditions [P1+P5+M3]) elicited a much earlier ($p < .0001$) and larger ($p < .0001$) POR than dissonant dyads (pooled conditions [m7+TT+m2]). Figure~\ref{fig:aefs}D depicts 99\,\% bootstrap confidence intervals for the POR amplitudes and latencies, pooled over hemispheres, in response to the experimental conditions; the activity pattern observed here also points to a close relationship between the degree of a dyad's consonance and the morphology of the respective POR.

    When pooling across conditions, we found a noticeable difference between the left and the right hemisphere in the POR amplitude ($p = .01$), but not in the POR latency ($p = .36$); also, the difference between the neuromagnetic responses to consonant or dissonant dyads did not significantly vary between hemispheres (latency: $p = .58$; amplitude: $p = .48$).

  \subsection{Neural mechanisms underlying the responses of auditory cortex to consonance and dissonance}

    The difference in POR latency in response to consonant and dissonant dyads in alGH suggests that consonance and dissonance are computed at relatively early stages of the cortical auditory hierarchy. We used our model of cortical pitch processing, designed to reproduce the neuromagnetic responses elicited by iterated rippled noises, to test this interpretation. If the differential responses to consonance and dissonance in alHG were intrinsic to pitch processing, we would expect our prospective mechanism to be able to reproduce this behaviour.

\begin{figure}[p!]
      \centering
      \includegraphics[width=1\textwidth]{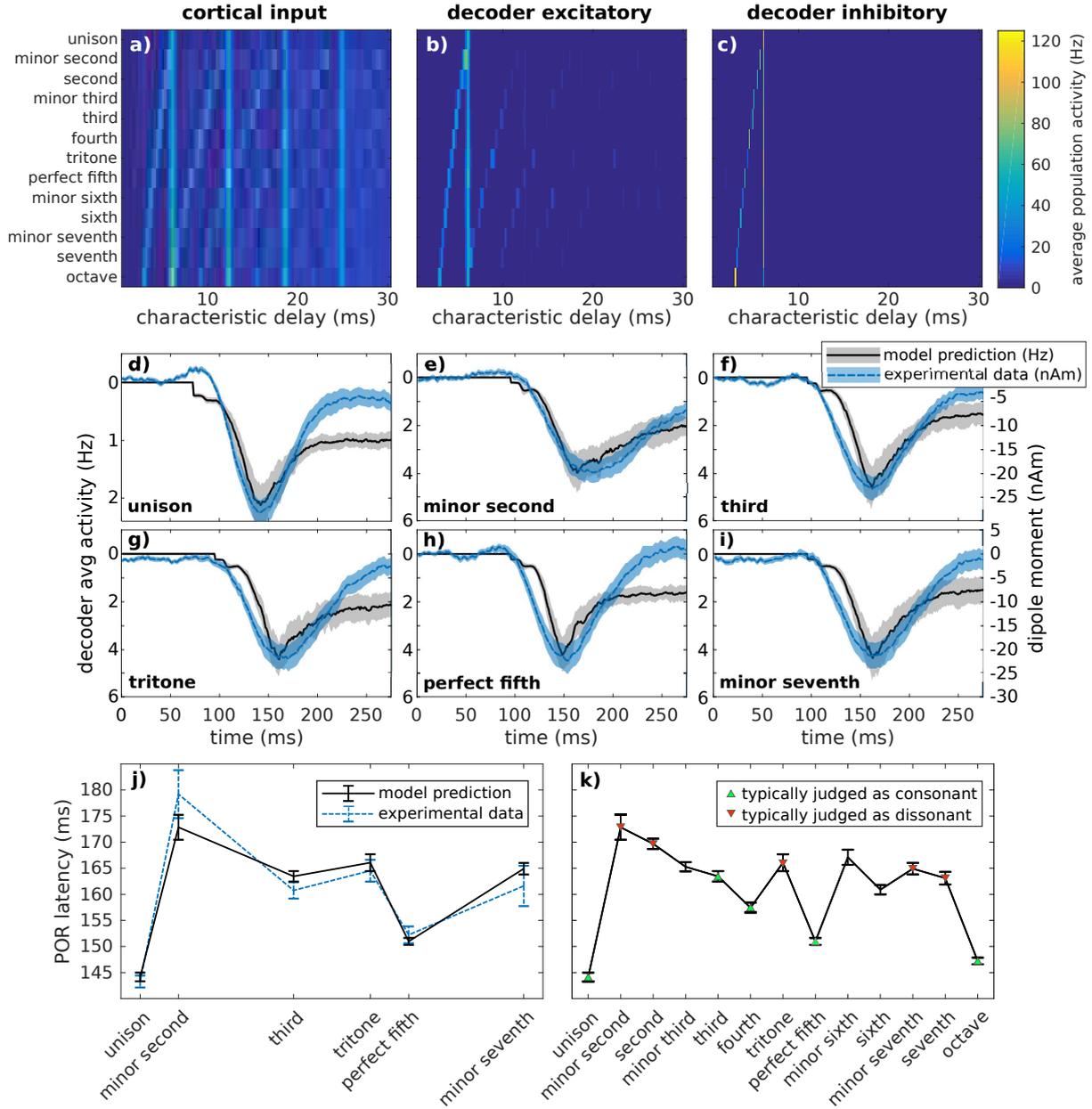}
      \caption{\textbf{Model responses to the IRN dyads used in the MEG experiment.}
      a)--c) Neural representation of the dyads at different levels of the model: (a) periodicity detectors, (b/c) excitatory/inhibitory ensembles in the decoder network; each row shows the activity elicited by each dyad. Excitatory and inhibitory ensembles in the sustainer are precisely correlated to the decoder-inhibitory heatmap. Note that, unlike in the perceptual results for simple IRNs, the neural activity representing the pitch of the second note shifts toward the left because dyads are arranged in ascending order. 
      d)--i) Examples of the collective excitatory activity at the decoder network (monotonically related to the equivalent dipole moment elicited by the network) in comparison with the elicited dipole moment measured during the experimentation in the neural generator of the POR. The scale of the field derived for the unison dyad was adjusted to account for the comparatively smaller effect on the network of the unison input, which effectively activates half of the populations than the other dyads.
      j) Latency predictions for IRN dyads compared with the experimental results reported in the previous section. 
      k) Latency predictions for the remaining dyads in the chromatic scale. 
      Strongly consonant dyads are represented with a green triangle, whilst strongly dissonant dyads are represented with a red triangle; dissonance was assessed according to Helmholtz~\cite{Helmholtz1863}, Table on Fig.~61. Dyads were generated using the same parameters as in the experimental procedures: the lower note pitch was set to 160\,Hz and the chromatic scale was generated using the \emph{just intonation} \cite{Helmholtz1863}.}
      \label{fig:mod:dyads}
    \end{figure}

    First, we verified that the model's was able to  provide a representation of the individual pitches of the two tones comprised in dyads; results are shown in Figure~\ref{fig:mod:dyads}a--c (see \supfigF for additional perceptual results obtained with different families of dyads). It should be emphasized that even phenomenological (i.e., non-mechanistic) models of pitch perception are generally unable to perform correct perceptual predictions for sounds with concurrent pitches (e.g.~\cite{Balaguer2009, Patterson1994a}; see~\cite{DeCheveigne2005b} for a review).

    Figure~\ref{fig:mod:dyads}j shows the latency predictions of the model are compared with the experimental data for the respective dyads. 
    Although the model predicted a slightly shorter POR latency for the semitone (m2) dyad than observed (see discussion), latency predictions faithfully reproduced the experimental trend; moreover, the differential response to consonant (P1, M3, P5) and dissonant (m2, TT, m7) dyads found in the MEG data was perfectly reproduced by the model (latency of P1 and P5 $<$ latency of dissonant dyads: $p<10^{-7}$, $U > 5150$; latency of M3 $<$ latency of m2: $p = 0.005$, $U = 4120$; latency of M3 $<$ latency of TT: $p = 0.63$, $U = 3571$; latency of M3 $<$ latency of m7: $p = 0.06$, $U = 3927$); according to pair-wise single-tailed Wilcoxon rank-sum tests performed over the results of $n = 60$ runs). The temporal dynamics of the dipole moment predicted by the model is shown in Figures~\ref{fig:mod:dyads}d--i.

    Last, we extended the POR latency predictions of the model to include all 13 dyads within the chromatic scale (see Figure~\ref{fig:mod:dyads}), and tested if the differential responses to consonance and dissonance were generalizable to additional dyads. Following Helmholtz~\cite{Helmholtz1863}, we considered an extended set of consonant dyads, including the octave (P8) and the perfect fourth (P4); and an extended set of dissonant dyads, including the major seventh (M7) and the major second (M2). Once again, consonant dyads produced longer latencies than dissonant dyads (latencies of P1, P4, P5 and P8 $<$ latencies of the extended set of dissonant dyads: $p<0.0003$, $U > 4290$; latency of M3 $<$ latency of M2: $p < 10^{-5}$, $U = 4497$; latency of M3 $<$ latency of M7: $p = 0.54$, $U = 3610$); according to pair-wise single-tailed Wilcoxon rank-sum tests performed over the results of 60 runs of the model). These results, fully in line with previous findings, reveal that the differential response of our model to consonance and dissonance in dyads is a general phenomenon caused by the fundamental relationships between the periodicities of the two dyad components. These analyses are extended to further families of dyads in \supfigG, yielding similar results.

    The model provides a mechanistic explanation of how the harmonic relationships between the components of the dyads modulate processing time. Consonant dyads consist of tones that share a larger number of lower harmonics than the ones in dissonant dyads. For instance, in the just intonation, the perfect fifth of a given fundamental shares one in every two harmonics with that fundamental, whilst only one in 16 harmonics are shared by a minor second and its fundamental. Our model suggests that cortical pitch processing is triggered by the joint activation of, at least, three periodicity detectors characterizing a specific harmonic series. Consonant dyads elicit a dramatically larger signal-to-noise ratio in the periodicity detectors tuned to their common harmonics, resulting in a collaborative effort towards pitch extraction that effectively speeds up processing dynamics (an animation of this process is depicted in \supfigB).

\section{Discussion}


    This work combines new theoretical and experimental methods to study how the auditory cortex representations of a sounds' pitch also generate the sensation of consonance and dissonance.

    First, we introduced a novel ensemble model of pitch designed to understand the neuromagnetic fields elicited in alHG during pitch processing. The model proposes a mechanistic explanation of the POR morphology and the dependence of its peak latency with the perceived pitch, a phenomenon that has been robustly observed for over two decades \cite{Krumbholz2003, Ritter2005, Seither2006}, yet remained poorly understood. Thereafter, we designed an MEG protocol to investigate whether the POR properties are influenced by the degree of consonance or dissonance, as elicited by different dyads from common Western music. Results revealed a strong correlation between the POR peak latency and the degree of dissonance elicited by each dyad, extending previous EEG results that have also reported a modulation of the POR amplitude by consonance \cite{Bidelman2013}.

    Last, we showed that our model, originally designed to explain pitch processing in IRN stimuli with a single pitch, can quantitatively account for the correlation between POR latency and the degree of consonance and dissonance. The model can explain the shorter POR latencies in response to consonant dyads as an effect of harmonic facilitation during pitch extraction. Combined, our results indicate that the neural mechanisms accounting for pitch processing show a differential response to consonant and dissonant dyads, suggesting that the sensation of consonance and dissonance might be elicited as a result of pitch processing in alHG.

  \subsection{The POR latency reflects pitch processing time}

    The dynamics of the decoder network proposes a new mechanistic interpretation for the POR latency in the sense that it might reflect the amount of time that is necessary for the network to stabilize in an unequivocal pitch (see Figure~\ref{fig:mod:dec}). Although an association between POR latency and processing time has been hypothesized previously in experiments (e.g.,~\cite{Alain1997, Krumbholz2003, Ritter2005, Seither2006}) and a in a model \cite{Balaguer2009}, a biophysical understanding of this phenomenon was still lacking. Our model identifies the magnitude of the POR with the instant in which the net inhibition at the decoder network exceeds the excitatory activity from the periodicity detectors; from a dynamic system perspective, this is equivalent to the instant in which the trajectory in the phase-space is unequivocally directed towards the attractor state dominated by the neural ensemble that is characterized by the perceived pitch (see animation in \supfigA).

    The model suggests that a robust perceptual decision concerning stimulus pitch is made after the cortical system identifies, typically, three peaks from the harmonic series of the stimulus' period in the representation of the periodicity detectors. This accounts for the relation between the POR latency and the stimulus period \cite{Krumbholz2003, Ritter2005}; moreover, it also may explain why pitch identification is only robust when the stimulus duration exceeds four times the pitch periodicity \cite{Krumbholz2003}. Previous studies have postulated that cortical pitch processing mechanisms must integrate along several period cycles in order to make a perceptual decision \cite{DeCheveigne2010, Gutschalk2004, Krumbholz2003}; however, a specific mechanism for such an integration has not been proposed up to now. 

    Moreover, since phase-locked activity is not robustly present above 50--200\,Hz in cortex \cite{Brugge2009}, integration along several repetition cycles was only possible in subcortical areas. The decoder network in our model takes advantage of the input harmonic representations provided by an autocorrelation model that does not require phase-locking to transmit information concerning several repetition cycles \cite{Meddis1997}, and thus provides a parsimonious solution to this problem.

  \subsection{Effect of consonance and dissonance on cortical processing time}

    Combined, our results suggest that cortical processing of dissonant dyads is slower than the processing of consonant dyads, in the sense that it requires a longer processing time. The model provides a physical rationale for this phenomenon: cortical extraction of consonance is based on the accumulation of activity in the columns with preferred periods characterizing the lower harmonics of the target sound; thus, concurrent pitch frequencies sharing common lower harmonics contribute to the build-up of each other's representation, thereby speeding up the stabilization of the network. Since consonant dyads are characterized by simpler frequency ratios, they comprise tones sharing a larger number of lower harmonics than dissonant dyads.

    Early phenomenological models based on Helmholtz's \emph{roughness} theory described dissonance as the beating sensation produced by tones with fundamental frequencies that were not harmonically related \cite{Helmholtz1863, Krueger1913, Plomp1965, Kameoka1969a, Sethares1993}. More recent explanations of consonance, based on pitch processing, have linked the regularity of the autocorrelation harmonic patterns elicited by dyads to their evoked consonance and dissonance percepts \cite{Bidelman2009, Bidelman2011, Bidelman2013, Ebeling2008, Tramo2006}. Thus, in one way or another, previous phenomenological models of consonance have consistently related perceived consonance with the amount of harmonics shared between the tones comprising the dyad. The present model confirms that assumption and introduces a potential explanation for its underlying biophysical rationale.

    Although our modelling results generally show a good fit with the data from the MEG experiment, the model prediction falls around 5\,ms short when explaining the POR latency evoked by the minor second dyad. This relatively small underestimation might result from the limited number of harmonics considered during the integration step in the decoder network: dissonant dyads, whose components do not share any common harmonic within the first three peaks of their harmonic series, present comparable processing times. An adaptive mechanism adjusting the number of harmonics required to trigger the decoding process according to the degree of overlap of the peaks in the input could potentially yield more accurate results. This adaptive mechanism would be necessary to explain how humans can differentiate dyads that differ in a quarter of a semitone.

    Our study did not evaluate, however, if the general (yet not universal \cite{Plantinga2014, Mcdermott2016}) association between consonance and pleasantness might be a consequence of differential processing in alHG. Future work should investigate whether this link could be due to different processing times, as described above, or whether it can be better explained by processes at higher levels of the auditory hierarchy that might be more sensitive to cultural and background modulations.

  \subsection{Experimental discussion and comparison with results of the literature}

    Our neuromagnetic findings concerning the POR morphology in response to consonant and dissonant dyads resemble and extend recent EEG data reported by Bidelman and Grall~\cite{Bidelman2014}, and by Proverbio et al. \cite{Proverbio2016}. Specifically, Bidelman and Grall~\cite{Bidelman2014} applied EEG in a smaller sample ($N$ = 9) of musically trained listeners and revealed a close relation between their subject's consonance/dissonance ratings and the morphology of the POR that was elicited by the respective dyads in alHG. In their study, the POR latency difference between consonant and dissonant dyads (cf. their figure\,4B) appears to have non-significant ($p = 0.22$) effect size, smaller than the results that were obtained in our study by means of MEG. 

    One reason for this might be that Bidelman and Grall~\cite{Bidelman2014} applied shorter IRN stimuli with a higher number of iterations, resulting in an increased saliency of the pitch percept; moreover, they employed a dichotic stimulation paradigm in which each ear was presented with only one dyad component, whereas in our experiment, sounds were delivered diotically to the listeners. On the other hand, the POR was found to originate from very similar locations in alHG, in both our MEG experiment and the work of Bidelman and Grall~\cite{Bidelman2014}; this is consistent with earlier fMRI \cite{Patterson2002} and intracranial \cite{Schoenwiesner2008} studies, as well as with other MEG studies that have linked subcomponents of the N100 wave to pitch processing (e.g., \cite{Gutschalk2004, Ritter2005, Andermann2014, Andermann2017}).

  \subsection{Relation to previous models of pitch processing}

    Previous studies have introduced a wide range of phenomenological models that were originally designed to predict the pitch of complex sounds (e.g.,~\cite{Licklider1951, Patterson1994a, Meddis1997, DeCheveigne1998, Balaguer2007, Balaguer2008, Balaguer2009}, see~\cite{DeCheveigne2005b} for a review). The correlation between pitch and cortical AEFs was previously addressed by the Auditory Image Model's \emph{buffer} \cite{Patterson1994a} and its derivative \cite{Gutschalk2007}; and the derivative of the activation in the population encoding the pitch in \cite{Balaguer2009, Tabas2016}. However, these models do not provide the biophysical mechanisms underlying the generation of the POR, or its latency dependence with pitch.

    Other models, designed to explain the precise biophysical mechanisms of pitch perception, were primarily focused on subcortical or early processing. Two of these models describe how neurons, mainly in subcortical nuclei, might process periodicities from the auditory nerve activity: Meddis and O'Mard's model \cite{Meddis2006} proposes a biophysical implementation of the summary autocorrelation function \cite{Licklider1951, Meddis1997}, based on the joint action of chopper neurons in the cochlear nucleus and coincidence detectors in the inferior colliculus.  More recently, Huang and Rinzel \cite{Huang2016} describe the neural implementation of a coincidence detector which is able to detect periodicities by comparing neural activity across different cochlear channels. Despite their mechanistic differences, both models present a spectral output comparable to that of the autocorrelation function \cite{Huang2016}. The model presented here is downstream, with respect to Meddis' and Huang's models, with its focus on explaining how pitch-related information is extracted from spectral patterns in alHG. Neurons implementing this transformation have been recently observed in the primary auditory cortex of marmosets \cite{Feng2017}.

  \subsection{Relation to previous models of sensory integration}

    The decoder network dynamics can be understood as an extension of the decision making model of the well-known winner-take-all ensemble competition \cite{Wang2002, Wong2006}: excitatory populations in the decoding network compete with each other, whilst the inhibitory ensemble that is arbitrating this selective competition is in the column sensitive to the fundamental (see the inhibitory connections in Figure~\ref{fig:mod:diagram}). Since competition is restricted to harmonically related populations, multiple fundamentals can be simultaneously processed by the decoder network in cases where the input presents concurrent pitch values (e.g., Figure~\ref{fig:mod:dyads}).

    The hierarchical structure of our model is inspired by a recent model that was designed to explain sensory integration in the medial temporal area of the macaque monkey \cite{Wimmer2015}. In that model, perceptual decisions are first transiently computed in a \emph{sensory circuit} that follows winner-take-all dynamics; after convergence, the decision then propagates to an \emph{integration circuit}, that further modulates the dynamics of the sensory circuit, thus ensuring stability until a significant change occurs in the input to the cortex. Similarly, once a pitch value has been extracted in the decoder network of our model, the activity of the winner column is reinforced by the sustainer network (rather than being repeatedly decoded) until a new change in the subcortical input triggers a new decoding process (see \supfigI).

    The sustaining strategy is also reminiscent of \emph{predictive coding} \cite{Auksztulewicz2017, Friston2005, Friston2009} and reversed hierarchical strategies \cite{Hochstein2002, Balaguer2009}, where top-down efferents convey expectations about the input (here: expectations regarding its harmonic structure), whereas bottom-up afferents convey prediction error (peaks not corresponding to the expected harmonic series) \cite{Rauss2013}. Additional top-down expectations could coexist at higher cognitive levels based on, for instance, prior knowledge, experience, or focused attention. Such biases could modulate the sustaining network by increasing the baseline activity of the inhibitory ensembles that characterize the target pitch values, thereby facilitating the extraction of that \emph{privileged} pitch in the decoder.


    To summarize, in this study we have introduced a novel model designed to understand the neural mechanisms of cortical pitch processing. The model proposes a mechanistic link between the latency of the POR subcomponent of the N100 wave and the processing time required for the system to achieve convergence, explaining the classical result that tones with a lower pitch elicit PORs with longer latencies. Moreover, our modelling and experimental results indicate that processing time varies with the degree of consonance in dyads, suggesting that the sensation of consonance and dissonance might directly stem from cortical pitch processing.

\section{Methods}

  \subsection{Experimentation}

    \subsubsection{Participants}

      Thirty-seven normal-hearing adults (22 female, 2 left-handed; mean age: 29.1 $\pm$ 8.3 years) participated in the experiment. None of the subjects reported any history of central or peripheral hearing impairments or any neurological or psychiatric disorders. The study and the experimental procedures were approved by the ethics committee of the Heidelberg University’s Medical School, and were conducted with written informed consent of each listener.

    \subsubsection{Stimuli}

      All stimuli were generated on-line using MATLAB 7.1 (The MathWorks, Inc., USA) and a sampling rate of 48000\,Hz. The basic stimulus was a 750\,ms long IRN segment, bandpass-filtered at 125--2000\,Hz, with eight iterations and gain for the delay-and-add filter $g_f = 1$. The delay of the IRN was varied between experimental conditions in an effort to build three consonant and three dissonant musical intervals, as classified by Western music theory. The delay of the lower note was always 6.25\,ms, corresponding to a pitch of 160\,Hz; the delay of the upper note was adjusted accordingly to form either a consonant dyad (unison, P1; perfect fifth, P5; major third, M3) or a dissonant dyad (tritone, TT; minor seventh, m7; minor second, m2). Table~\ref{tab:expCond} presents an overview of the six experimental conditions.  

      In order to separate the dyad-specific neuromagnetic responses from the cortical activity associated with the onset of sound energy \cite{Biermann2000, Gutschalk2002}, each IRN dyad was preceded by a 750\,ms long, energy-balanced noise segment (bandpass-filtered at 125--2000\,Hz). There were 10\,ms Hanning windows at stimulus onset and offset; moreover, between the first (noise) and the second (IRN) segment of a stimulus, signals were cross-faded for a duration of 10\,ms in an effort to avoid discontinuous waveforms. The overall stimulation level was set to 80\,dB SPL.

      \begin{table}
        \centering
        \begin{tabular}{ccccc}
	\toprule
	\textbf{musical interval}   & freq. ratio & freq. (rounded, Hz) & \# of semitones & consonance percept. \\
	\midrule
	\textbf{unison} (P1)        &    1:1      &     160 / 160       &         0       &       PC            \\
	\textbf{perfect fifth} (P5) &    3:2      &     160 / 240       &         7       &       PC            \\
	\textbf{major third} (M3)   &    5:4      &     160 / 200       &         4       &       IC            \\
	\textbf{tritone} (TT)       &    45:32    &     160 / 225       &         6       &       D             \\
	\textbf{minor seventh} (m7) &    16:9     &     160 / 284       &        10       &       D             \\
	\textbf{minor second} (m2)  &    16:15    &     160 / 171       &         1       &       D             \\ 
	\bottomrule
\end{tabular}
        \caption{Overview of the experimental conditions. Dyads are listed in descending consonance order, and are categorized as Perfect consonant (PC), imperfect consonant (IC) or dissonant (D) according to Western music theory and empirical results \cite{Schwartz2003, Itoh2010}.}
        \label{tab:expCond}
      \end{table}

    \subsubsection{Data acquisition and processing}

      Gradients of the magnetic field were acquired with a Neuromag-122 whole-head MEG system (Elekta Neuromag Oy, Helsinki, Finland) inside a magnetically shielded room (IMEDCO, H\"{a}gendorf, Switzerland). Raw data were acquired at a sampling rate of 1000\,Hz and low-pass filtered at 330\,Hz. Prior to the recordings, the nasion, two pre-auricular points and 32 surface points were measured as anatomical landmarks, individually for each participant, using a Polhemus 3D-Space Isotrack2 systems. In an effort to keep vigilance stable, participants watched a silent movie of their own choice during data acquisition, and they were asked to direct their attention to the movie and ignore the sounds in the earphones. The IRN dyads were delivered to the subjects via Etymotic Research (ER3) earphones with 90\,cm plastic tubes and foam earpieces. Sounds were presented using a 24-bit sound card (RME ADI 8DS AD/DA interface), an attenuator (Tucker-Davis Technologies PA-5) and a headphone buffer (Tucker-Davis Technologies HB-7). 250 sweeps per stimulus condition were played during the MEG recording, diotically and in pseudo-randomized order. The inter-stimulus interval was 1000\,ms. The total duration of the measurement was 62 minutes.

    \subsubsection{Data analysis}

      Data were analyzed off-line using the BESA 5.2 software package (BESA GmbH, Germany) with a spherical head model and a homogeneous volume conductor. After visual inspection of the raw data, noisy channels and sweeps with amplitudes greater than 8000\,fT/cm or gradients exceeding 800\,fT/cm/ms were excluded from further analyses. About 235 sweeps per subject and condition remained after artifact rejection; they were averaged, trigger-synchronously, in the epoch from 500\,ms before to 3000\,ms after stimulus onset. The baseline was defined as the average level in the interval of -100\,ms to 0\,ms, relative to stimulus onset.

      After pre-processing, we applied spatio-temporal source models \cite{Scherg1989, Scherg1990, Scherg1991} in BESA, in an effort to study the POR component in response to the second stimulus segment, i.e., at the transition from noise to IRN dyads. In this source localization approach, the intracortical sources of the activity observed at the scalp are modeled as equivalent current dipoles, and their spatial position and orientation is varied iteratively until a maximum amount of variance is explained in the scalp data. The source model includes both the spatial information for each dipole, and its physiological activity across time (source waveform). We calculated source models with one dipole per hemisphere for the POR component in the second stimulus segment. Dipole fits were based on pooled conditions [P1+P5+M3+TT+m7+m2]. The fitting interval covered about 30\,ms around its peak, and MEG data were zero-phase filtered 2--20\,Hz.

      Individual fits at the AEF components were successful for 36 subjects. In ten participants we included a symmetry constraint in the model to stabilize the individual dipole fits. One participant failed to show stable fits in the dipole model, and was excluded from subsequent analyses. Aside from symmetry, no further constraints were made concerning the orientation and location of the dipoles. The average maximum of explained variance within the fitting window was 64.1\% (SD: 18.9) for the POR dipole model. After fitting, this dipole model was used as spatio-temporal filter, i.e., the source waveforms corresponding to the model were extracted separately for each condition and each subject. Finally, the source waveforms were exported from BESA to MATLAB for statistical analysis.

      The statistical evaluation of the MEG source waveforms was conducted using the bootstrap method \cite{Efron1993}. Here, the distribution of a test statistic is approximated by repeated random drawing, with replacement, from the original dataset; based on the resulting bootstrap distribution, confidence intervals can then be derived for that test statistic. Contrary to most standard techniques, the bootstrap method is well-suited for neurophysiological data where peaks cannot be clearly identified for each participant in every condition. Prior to statistical analyses, each source waveform of the POR model was adjusted to the baseline calculated as the average of the last 100\,ms before the transition.

  \subsection{Modelling}

    \subsubsection{Peripheral model and periodicity detectors}

      Neural activity at the auditory nerve was simulated using a recent biophysically realistic model of the auditory periphery \cite{Zilany2009, Zilany2014}. Peripheral parameters were chosen as in~\cite{Meddis2006}, considering 40 cochlear channels with centre frequencies between 125\,Hz and 10\,kHz.

      Periodicity detectors were modeled according to the summarized autocorrelation function (SACF) of the auditory nerve activity \cite{Meddis1997, Meddis2006, Balaguer2008}. This highly idealized model yields a harmonic neural representation of pitch-related information (see Figure~\ref{fig:mod:diagram}e), often connected to subcortical processing. The SACF was chosen for its comparably low computational complexity, but more detailed biophysical models yielding similar representations (e.g.~\cite{Meddis2006, Huang2016}) should produce comparable results.

      The SACF used here follows the same formulation as the first stage in the cascade autocorrelation model \cite{Balaguer2008}. The $n$th component $A_n(t)$ of the SACF represents a measure of the regularity of the auditory nerve activity with respect to a fixed period $\delta t_n$. The model considers $N = 250$ components $A_n(t)$ with characteristic periods uniformly spaced between $\delta t_1 = 0.5\,$ms, a conservative estimation of the phase-locking limit of the auditory nerve \cite{Bendor2012a}, up to the lower limit of melodic pitch, $\delta t_N = 30$\,ms \cite{Pressnitzer2001}.

      The output is further regularized through a procedure $A_n(t) \rightarrow \hat{A}_n(t)$ that reduces the dependence of the SACF with stimulus intensity level and minimizes signal-to-noise variations within sounds with the same pitch but different timbre. The regularization procedure makes use of the principles of neural normalization \cite{Carandini2012} (see details in Supplementary Methods).

    \subsubsection{Ensemble dynamics}

      Neural ensembles follow mean-field dynamics characterised by their instantaneous firing rate $H^e_n(t)$ (excitatory) and $H^i_n(t)$ (inhibitory), at each cortical column $n$. Evolution dynamics were adapted from~\cite{Wong2006}:

      \begin{equation}
        \tau^{\text{pop}} \, \dot{H}^{e,i}_n(t) = - H^{e,i}_n(t) + \phi^{e,i}(I^{e,i}_n(t))
        \label{eq:Hei}
      \end{equation}

      \noindent with transfer functions $\phi^{e,i}(I^{e,i}_n(t))$ \cite{Wong2006}:

      \begin{equation}
        \phi^{e,i}(I) = \frac{a^{e,i} I - b^{e,i}}{1 - e^{-d^{e,i} (a^{e,i} I - b^{e,i})}}
        \label{eq:transfer}
      \end{equation}

      Realistic parameters of the excitatory and inhibitory transfer functions ($a^e$, $b^e$ and $d^e$ for the excitatory; $a^i$, $b^i$ and $d^i$ for the inhibitory) were taken from the literature \cite{Brunel2001, Wong2006}. The total synaptic inputs $I^e_n(t)$ and $I^i_n(t)$ are defined below. Numerically simulations were performed using the Euler's method with a time step $\Delta t = 1$\,ms.

      Dynamics of excitatory and inhibitory ensembles at the \emph{decoder} and \emph{sustainer} networks follow the same formulation. In order to differentiate between the two cortical networks, we use $H^{e,i}_n(t)$ and $I^{e,i}_n(t)$ to characterize populations and synaptic inputs of the decoder layer, and $\hat{H}^{e, i}_n(t)$ and $\hat{I}^{e,i}_n(t)$ for the  populations and synaptic inputs of the sustainer layer. Population effective time constants $\tau^{\text{pop}}$ are adaptive and depend on the activity of the population \cite{Ostojic2011, Gerstner2014}:

      \begin{equation}
        \tau^{\text{pop}}(H(t)) = \tau^{\text{pop}}_0 \, \Delta_T \frac{\phi'(I(t))}{H(t)}
        \label{eq:taupop}
      \end{equation}

      \noindent where $\phi'(I(t))$ is the slope of the transfer function (see Equation~\ref{eq:transfer}) and $\Delta_T$ is the sharpness of the action potential initiation. The mean-field dynamics of the populations in our model was based on a LIF neuron \cite{Wong2006} that approximates the action potential initiation as instantaneous \cite{Fourcaud2003}; thus, we used a small $\Delta_T \ll 1 = 0.05$\,mV.

    \subsubsection{Synaptic dynamics}

      Ensemble connectivity is mediated through realistic AMPA, NMDA and GABA$_\text{A}$ synapses \cite{Brunel2001, Wang2002, Wong2006, Deco2013}. Synaptic dynamics were modelled according to Brunel and Wang formulation \cite{Brunel2001}:

      \begin{eqnarray}
        \dot{S}_n^{\text{AMPA}}(t) & = &
           - \frac{S_n^{\text{AMPA}}(t)}{\tau_{\text{AMPA}}}
           + H_n^e(t) + \sigma \nu_n(t)             \label{eq:Sampa} \\
        \dot{S}_n^{\text{GABA}}(t) & = &
           - \frac{S_n^{\text{GABA}}(t)}{\tau_{\text{GABA}}}
           + H_n^i(t) + \sigma \nu_n(t)             \label{eq:Sgaba} \\
        \dot{S}_n^{\text{NMDA}}(t) & = &
           - \frac{S_n^{\text{NMDA}}(t)}{\tau_{\text{NMDA}}}
           + \gamma \left(1 - S_{\text{NMDA}}(t)\right) H_n^e(t)
           + \sigma \nu_n(t)                        \label{eq:Snmda}
      \end{eqnarray}

      NMDA time constant was set to $\tau_{\text{NMDA}} = 30\,$ms; GABA and AMPA time constants $\tau_{\text{GABA}} = 2\,$ms and $\tau_{\text{AMPA}} = 5\,$ms, and the coupling parameter $\gamma = 0.641$, were all taken from the literature \cite{Wong2006, Brunel2001}. Gating variables at the sustainer and decoder layers $\hat{S}_n^{\text{NMDA, AMPA, GABA}}(t)$, $\hat{H}^{e,i}_n(t)$ follow similar dynamics.

    \subsubsection{Synaptic inputs}

      Total synaptic inputs to populations $I^{i,e}_n(t)$ and $\hat{I}^{i,e}_n(t)$ consist of three different contributions: internal input $I_{\text{int}}$, accounting for inputs from populations within the same network; external input $I_{\text{ext}}$, exerted by sources from other networks; and a constant input drive $I_0$:

      \begin{eqnarray}
        I^{i,e}_n(t)       & = & I^{i,e}_{n, \text{int}}(t) +
                                 I^{i,e}_{n, \text{ext}}(t) +
                                 I^{i,e}_{n, 0}(t)
                                              \label{eq:totalI:dec} \\
        \hat{I}^{i,e}_n(t) & = & \hat{I}^{i,e}_{n, \text{int}}(t) +
                                 \hat{I}^{i,e}_{n, \text{ext}}(t) +
                                 \hat{I}^{i,e}_{n, 0}(t)
                                              \label{eq:totalI:sus}
      \end{eqnarray}

      \paragraph{Internal input} Connectivity weights between any two ensembles in the decoder network are provided by the matrices $C^{ee}, C^{ei}, C^{ie}, C^{ii}$. $C^{ei}$ and $C^{ie}$. $C^{ei}$ and $C^{ie}$ present a harmonic structure inspired in connectivity patterns reported in the mammal auditory cortex (see \cite{Wang2013} for a review); this matrices are plotted in Figure~\ref{fig:mod:diagram}d. $C^{ee}$ is the identity matrix, and $C^{ii}$ has a similar diagonal structure: $C^{ii}_{\alpha\beta} = (1 - c^{ie}_0) \delta_{\alpha\beta} + c^{ie}_0$, where $c^{ie}_0$ is the baseline inhibitory weight $c^{ie}_0 = 0.1$ and $\delta_{\alpha\beta}$ is the Kronecker delta.

      Altogether, the internal inputs at the decoder $I_{\text{int}}(t)$ are defined as follows:

      \begin{eqnarray}
        I^e_{n, \text{int}}(t)       & = & \sum_k C^{ee}_{nk} \left(
                J^{ee}_{\text{NMDA}} \, S_k^{\text{NMDA}}(t) +
                J^{ee}_{\text{AMPA}} \, S_k^{\text{AMPA}}(t)
                  \right) - \sum_k C^{ie}_{nk}
                J^{ie}_{\text{GABA}} \, S_k^{\text{GABA}}(t)
                                              \label{eq:Iint:edec} \\
        I^i_{n, \text{int}}(t)       & = & \sum_k C^{ei}_{nk} \left(
                J^{ie}_{\text{NMDA}} \, S_k^{\text{NMDA}}(t) +
                J^{ei}_{\text{AMPA}} \, S_k^{\text{AMPA}}(t)
                  \right) - \sum_k C^{ii}_{nk}
                J^{ii}_{\text{GABA}} \, S_k^{\text{GABA}}(t)
                                              \label{eq:Iint:idec}
      \end{eqnarray}

      Ensembles ni the sustainer network only communicate internally with ensembles within the same block:

      \begin{eqnarray}
        \hat{I}^e_{n, \text{int}}(t) & = &
                \hat{J}^{ee}_{\text{NMDA}} \, \hat{S}_n^{\text{NMDA}}(t) +
                \hat{J}^{ee}_{\text{AMPA}} \, \hat{S}_n^{\text{AMPA}}(t) -
                \hat{J}^{ie}_{\text{GABA}} \, \hat{S}_n^{\text{GABA}}(t)
                                              \label{eq:Iint:esus} \\
        \hat{I}^i_{n, \text{int}}(t) & = &
                \hat{J}^{ei}_{\text{NMDA}} \, \hat{S}_n^{\text{NMDA}}(t) +
                \hat{J}^{ei}_{\text{AMPA}} \, \hat{S}_n^{\text{AMPA}}(t) -
                \hat{J}^{ii}_{\text{GABA}} \, \hat{S}_n^{\text{GABA}}(t)
                                              \label{eq:Iint:isus}
      \end{eqnarray}

      \noindent Conductivities $J_{\text{NMDA}, \text{AMPA}, \text{GABA}}$ and $\hat{J}_{\text{NMDA}, \text{AMPA}, \text{GABA}}$ (see Table~\ref{tab:pars}) were initialized to typical values in the literature $J \simeq 0.15\,$nA \cite{Wong2006}, and fine-tuned within a range of realistic values to ensure the convergence of the ensembles activity to match perceptual results in iterated rippled noises \cite{Krumbholz2003}. Model's final parameters are listed in Table~\ref{tab:pars}.

      \begin{table}[p]
        \centering
        \begin{tabular}{c r @{\,} l c}
  \toprule
            parameter          &      \multicolumn{2}{c}{value}    &        source         \\  
  \midrule
        $J^{th}_{\text{AMPA}}$ & $2.6             $ & nA           &        fitted         \\  
         $J^{e}_{\text{NMDA}}$ & $0.45            $ & nA           &        fitted         \\  
   $\hat{J}^{a}_{\text{GABA}}$ & $0.45            $ & nA           &        fitted         \\  
   $\hat{J}^{a}_{\text{AMPA}}$ & $0.35            $ & nA           &        fitted         \\  
        $J^{ee}_{\text{NMDA}}$ & $0.14            $ & nA           &        fitted         \\  
        $J^{ee}_{\text{AMPA}}$ & $9.9\times10^{-4}$ & nA           &    \cite{Wong2006}    \\  
        $J^{ei}_{\text{NMDA}}$ & $0.20            $ & nA           &        fitted         \\  
        $J^{ei}_{\text{AMPA}}$ & $6.5\times10^{-5}$ & nA           &    \cite{Wong2006}    \\  
        $J^{ie}_{\text{GABA}}$ & $0.55            $ & nA           &        fitted         \\  
        $J^{ii}_{\text{GABA}}$ & $0.11            $ & nA           &        fitted         \\  
  $\hat{J}^{ee}_{\text{NMDA}}$ & $0.25            $ & nA           &        fitted         \\  
  $\hat{J}^{ee}_{\text{AMPA}}$ & $9.9\times10^{-4}$ & nA           &    \cite{Wong2006}    \\  
  $\hat{J}^{ei}_{\text{NMDA}}$ & $0.00            $ & nA           &        fitted         \\  
  $\hat{J}^{ei}_{\text{AMPA}}$ & $9.9\times10^{-4}$ & nA           &    \cite{Wong2006}    \\  
  $\hat{J}^{ie}_{\text{GABA}}$ & $0.80            $ & nA           &        fitted         \\  
  $\hat{J}^{ii}_{\text{GABA}}$ & $0.00            $ & nA           &        fitted         \\  
                    $c^{ie}_0$ & $0.1             $ &              &        fitted         \\  
                      $\gamma$ & $0.641           $ &              &   \cite{Brunel2001}   \\  
                         $a^e$ & $310             $ & (VnC)$^{-1}$ &    \cite{Wong2006}    \\  
                         $b^e$ & $125             $ & Hz           &    \cite{Wong2006}    \\  
                         $d^e$ & $0.16            $ & s            &    \cite{Wong2006}    \\  
                         $a^i$ & $615             $ & (VnC)$^{-1}$ &    \cite{Wong2006}    \\  
                         $b^i$ & $177             $ & Hz           &    \cite{Wong2006}    \\  
                         $d^i$ & $0.087           $ & s            &    \cite{Wong2006}    \\  
                       $I^e_0$ & $0.315           $ & nA           &        fitted         \\  
                       $I^i_0$ & $0.15            $ & nA           &        fitted         \\  
                 $\hat{I}^e_0$ & $0.26            $ & nA           &        fitted         \\  
                 $\hat{I}^i_0$ & $0.18            $ & nA           &        fitted         \\  
          $\tau_{\text{AMPA}}$ & $2               $ & ms           &   \cite{Brunel2001}   \\  
          $\tau_{\text{GABA}}$ & $5               $ & ms           &   \cite{Brunel2001}   \\  
          $\tau_{\text{NMDA}}$ & $30              $ & ms           &         fixed         \\  
           $\tau^{\text{pop}}$ & $10              $ & ms           &   \cite{Ostojic2011}  \\  
                    $\Delta_T$ & $0.05            $ & mV           &         fixed         \\  
                      $\sigma$ & $0.007           $ & nA           &    \cite{Wong2006}    \\  
          $\tau^{\text{adap}}$ & $100             $ & ms           &  \cite{Gerstner2002}  \\  
                   $\alpha = $ & $3\times10^{-6}  $ & nA           &        fitted         \\  
                               &                   &              &                       \\  
  \bottomrule  
\end{tabular}

        \caption{\textbf{Values for the parameters used in the cortical model} source in the literature.}{The Last column specifies the source of the parameter value; entries with the label \emph{fitted} were tuned according to the indications described in the main text; entries with the label \emph{fixed} were selected to a fixed value according to theoretical considerations (see main text).}
        \label{tab:pars}
      \end{table}

      \paragraph{External input} Excitatory ensembles in the decoder network receive bottom-up input $\hat{A}_n(t)$ via AMPA-driven synapses, according to previous studies in perceptual integration \cite{Wong2006}:

      \begin{equation}
        I^e_{n, \text{ext}}(t) = J^{th}_{\text{AMPA}} \,
                       S_n^{th, \text{AMPA}}(t)   \label{eq:Iext:esus}
      \end{equation}

      \noindent The conductivity $J^{th}_{\text{AMPA}}$ was adjusted to ensure a smooth and robust propagation of the activity in the periodicity detectors to the decoder's excitatory populations. The corresponding gating variables $S_n^{th, \text{AMPA}}(t)$ follow AMPA-like dynamics:

      \begin{equation}
        \dot{S}_n^{th, \text{AMPA}}(t) =- \frac{S_n^{th, \text{AMPA}}(t)}
               {\tau_{\text{AMPA}}} + A_n(t)       \label{eq:SampaTh}
      \end{equation}

      Inhibitory ensembles in the decoder receive efferent external input from the sustainer network. Top-down excitatory processes in cortex are typically dominated by NMDA dynamics \cite{Friston2005}; thus, efferent AMPA synapses were not considered:

      \begin{equation}
        I^i_{n, \text{ext}}(t) = J^{e}_{\text{NMDA}} \,
              \hat{S}_n^{th, \text{NMDA}}(t) \label{eq:isus}
      \end{equation}

      The efferent conductivity $J^{e}_{\text{NMDA}}$ (Table~\ref{tab:pars}) was tuned to enable the top-down enhancement of the inhibitory ensembles at the decoder once the lower harmonics that have been inhibited after the decoding process (see \emph{Dynamics of the decoder network} in Results).

      Sustainer's external inputs are sourced in the decoder network, driven by inhibitory GABAergic and excitatory AMPAergic synapses \cite{Wong2006, Friston2005}:

      \begin{eqnarray}
        \hat{I}^e_{n, \text{ext}}(t) & = &
                   \hat{J}^{a}_{\text{AMPA}} \, S_n^{\text{AMPA}}(t)
                                                \label{eq:hIext:esus} \\
        \hat{I}^i_{n, \text{ext}}(t) & = &
                   \hat{J}^{a}_{\text{GABA}} \, S_n^{\text{GABA}}(t)
                                                 \label{eq:hIext:isus}
      \end{eqnarray}

      Afferent conductivities $\hat{J}^{a}_{\text{AMPA, GABA}}$ (Table~\ref{tab:pars}) were set to make the sustainer both sensitive to decoded decisions, yet robust to spurious activations.

      \paragraph{Constant input drive} Constant inputs to the decoder $I^e_{n, 0}(t) = I^e_0$ and  $I^i_{n, 0}(t) = I^i_0$ (Table~\ref{tab:pars}) were selected to enable the system to be reactive to external input, yet silent in absence of a significant input. An additional constant drive $I^{\text{sus}}_0 = 0.24\,$nA was applied to the populations at the sustainer (see \emph{Dynamics of the sustainer network} in Results).

    \subsubsection{Derivation of the evoked fields}

      Assuming that all microcolumns within each of the two cortical networks present similar orientations, the total dipolar moment representing the neuromagnetic field elicit by each network is monotonicallt related to the collective excitatory activity in the network \cite{Kiebel2008, Kerr2008, Bruyns2016}:

      \begin{equation}
        m(t) = \sum_n H^e_n(t + \Delta t_{\text{subcort}})
      \end{equation}

      The subcortical delay $\Delta t$ accounts for the time elapsed from tone onset until the signal first arrives in the decoder network. This delay reflects not only propagation time, but also processing time of secondary processes like the regularization of the output of the periodicity detectors.

      The delay was fixed to $\Delta t = 75\,$ms, according to the experimentally observed latency of the POR elicited by an IRN with a delay of 8\,ms. We used a longer $\Delta t^{\text{dyads}} = 95\,$ms in dyads to compensate for a systematic 20\,ms delay of the experimental observations, most likely due to the different rescaling factors used for the regularized SACF of simple tones and dyads (see Supplementary Methods for details).

\bibliography{bib}{}
\bibliographystyle{ieeetr}

\end{document}